\documentclass[preprints,article,accept,moreauthors,pdftex]{Definitions/mdpi}

\firstpage{1}
\pubvolume{7}
\issuenum{1}
\articlenumber{15}
\pubyear{2021}
\copyrightyear{2021}
\datereceived{1 December 2020}
\dateaccepted{10 January 2021}
\datepublished{14 January 2021}
\hreflink{https://doi.org/10.3390/\\universe7010015} 

\pdfoutput=1

\usepackage{bm}

\usepackage{graphicx}
\usepackage{longtable}
\usepackage{subfigure}
\usepackage{threeparttable}

\Title{Intra-Day Variability Observations of Two Dozens of Blazars at 4.8 GHz}

\TitleCitation{Intra-Day Variability Observations of Two Dozens of Blazars at 4.8 GHz}

\Author{{Xiang Liu} 
$^{1,2,3,}$*\href{https://orcid.org/0000-0001-9815-2579}{\orcidicon}, {Xin} Wang $^{1,4}$\href{https://orcid.org/0000-0001-8221-9601}{\orcidicon}, {Ning} Chang $^{1,4}$\href{https://orcid.org/0000-0002-8684-7303}{\orcidicon}, {Jun} Liu $^{5}$, {Lang} Cui $^{1,2}$\href{https://orcid.org/0000-0003-0721-5509}{\orcidicon}, {Xiaofeng} Yang $^{1,2}$ \mbox{and {Thomas~P.} Krichbaum $^{5}$}}

\AuthorNames{X. Liu, X. Wang, N. Chang, J. Liu, L. Cui, X. Yang and T.~P. Krichbaum}

\AuthorCitation{Liu, X.; Wang, X.; Chang,~N.; Liu, J.; Cui, L.; Yang, X.; Krichbaum, T.P.}

\address{%
$^{1}$ \quad Xinjiang Astronomical Observatory, Chinese Academy of Sciences, 150 Science 1-Street, \mbox{Urumqi 830011, China;} wangxin2019@xao.ac.cn (X.W.); changning@xao.ac.cn (N.C.); \mbox{cuilang@xao.ac.cn (L.C.);} yangxiaofeng@xao.ac.cn (X.Y.)\\
$^{2}$ \quad Key Laboratory of Radio Astronomy, Chinese Academy of Sciences, Nanjing 210008, China\\
$^{3}$ \quad Physics and Astronomy Department, Qiannan Normal University for Nationalities, Duyun 558000, China\\
$^{4}$ \quad School of Astronomy and Space Science, University of Chinese Academy of Sciences, Beijing 100049, China\\
$^{5}$ \quad Max-Plank-Institut f\"ur Radioastronomie, Auf dem H\"ugel 69, 53121 Bonn, Germany; \mbox{jliu@mpifr-bonn.mpg.de (J.L.);} tkrichbaum@mpifr-bonn.mpg.de (T.P.K.)}

\corres{Correspondence: liux@xao.ac.cn}

\abstract{Two dozens of radio loud active galactic nuclei (AGNs) have been observed with Urumqi 25 m radio telescope in order to search for intra-day variability (IDV). The target sources are blazars (namely flat spectrum radio quasars and BL Lac objects) which are mostly selected from the observing list of RadioAstron AGN monitoring campaigns. The observations were carried out at 4.8 GHz in two sessions of 8--12 February 2014 and 7--9 March respectively. We report the data reduction and the first results of observations. The results show that the majority of the blazars exhibit IDV in 99.9\% confidence level, some of them show quite strong IDV. We find the strong IDV of blazar 1357 + 769 for the first time. The IDV at centimeter-wavelength is believed to be predominately caused by the scintillation of blazar emission through the local interstellar medium in a few hundreds parsecs away from Sun. No significant correlation between the IDV strength and either redshift or Galactic latitude is found in our sample. The IDV timescale along with source structure and brightness temperature analysis will be presented in a forthcoming paper.}


\begin{document}


\section{Introduction}\label{sec:intro}
The intra-day variability (IDV) of active galactic nuclei (AGNs) at centimeter wavelengths was first discovered in 1980s~\cite{Witzel1986, Heeschen1987}. Since then, more observations have been carried out, and~it turns out that the IDV occurs only in flat spectrum radio sources (FSRSs) which are core-dominated AGNs, with~a detection rate of 25--50\% in the FSRSs~\cite{Quirrenbach1992, Lovell2008} and with a higher rate of $\sim$60\% in bright {\it {Fermi}} 
blazars~\cite{Liux2012a}.

For a long time, it is controversial if the origin of the radio IDV at centimeter wavelengths, is AGN-intrinsic variations or is due to scattering-induced variations of AGN signals by ionized interstellar medium in our Galaxy~\cite{WagnerWitzel1995, Jauncey2016}. If~the IDV is AGN-intrinsic, the~timescale of a day or less will imply very high brightness temperatures of the jet-emitting region, several magnitudes (typically {\mbox{$10^{3-5}$}} 
times) greater than the inverse-Compton limit of \(T_{\rm IC}\sim 10^{11.5}\) K~\cite{Kellermann1969} or the equipartition temperature of \(T_{\rm eq}=5\times 10^{10}=10^{10.7}\) K~\cite{Readhead1994}. On~the other hand, the~brightness temperatures ($T_{b}$) measured from the RadioAstron (RA) space VLBI (SVLBI) observations, are 10--100 times larger than the IC limit, in~a significant fraction of blazars~\cite{Lobanov2015, Kovalev2016, Kravchenko2020, Zensus2020}. The~$T_{b}$ can be boosted by the Doppler beaming effect when the jet of blazar points closely to our line of sight, i.e.,~$T_{b}^{obs}=\delta T_{b}^{int}$, where $T_{b}^{obs}$ is the observed brightness temperature and $T_{b}^{int}$ is brightness temperature in source frame. However, the~Doppler factor $\delta$ is $<$60 for blazars observed from ground-based VLBI observations~\cite{Cohen2007, Liodakis2017, Liodakis2018} and mostly $\delta \le10$ \cite{Kovalev2016}, which is not high enough to account for the $T_{b}$ excess in the space VLBI observations, implying that the IC limit might be violated in some of the blazars. Nevertheless, the~apparent `brightness temperature' implied from the IDV, as~mentioned above, is much higher than the $T_{b}$ of blazars measured with the RA space~baselines.

\textls[-5]{Today, it becomes clear that the IDV at centimeter wavelengths is not AGN-intrinsic, but caused by the interstellar scintillation (ISS) effect in our Galaxy~\cite{Jauncey2020}.~The evidence is from the annual modulations of IDV timescales discovered in a couple of IDV sources that had been monitored for years, which can be well fitted by the Earth orbiting around the Sun and annually across the scintillation pattern (e.g., \cite{Dennett-Thorpe2003, Bignall2006, Liux2012b}, see a summary on this in~\cite{Bignall2019}).}

\textls[-5]{Furthermore, the flux fluctuations caused by the ISS-induced IDV, may have contributed partly to the RA SVLBI observed {$T_b$} excess (over the IC limit) of blazars~\cite{Liuj2018}. So it is important to find and study the IDV from the blazars in the RA SVLBI observations~\cite{Kovalev2020}.
In order to search for IDV in the RA sources in parallel with the RA SVLBI sessions, we carried out the IDV observations of the RA sources (plus a few blazars from other samples) with the Urumqi 25~m radio telescope in Xinjiang Astronomical Observatory. In~this paper, we report {the data reduction and the first results of the observations performed in search of IDV.}}

\section{Observations and Data~Reduction}

The IDV observations are carried out in {8 February 2014} 
05:12--{12 February 2014} 03:21 (94 h in total) and {7 March 2014} 03:52--{9 March 2014} 17:53 (62 h in total) with the Urumqi 25~m radio telescope. Most of the blazars are selected from the 2014 February and March sessions of RadioAstron space VLBI program~\cite{Kovalev2020}, with~the declination $>+10^{\circ}$, plus several additional blazars selected from previous IDV observations. The~observations were done with the 25~m telescope at the central frequency of 4.8 GHz with the bandwidth of 600~MHz, the lowest system temperature of 24~K, and~the antenna sensitivity of $\sim$0.12~K/Jy.

The observations were made with the cross-scans mode consisting of 4 + 4 sub-scans in azimuth and elevation over the source position. After~initial calibration of raw data, the~intensity profile of each sub-scan is fitted with a baseline and a Gaussian function after subtracting the baseline, and~then the data are averaged in 4 azimuth sub-scans and \mbox{4 elevation} sub-scans respectively. The~averaged azimuth amplitude is corrected with the pointing error in elevation direction, and~vice verse. The~antenna gain curve is applied for the data averaged from the azimuth and elevation scans, the~gain curve (including the air-mass effect) is measured from the secondary calibrators in the same observation. A~systematic time-dependent effect which is derived from the secondary calibrators is applied. Finally, the~amplitude is converted to the absolute flux density with the average scale of the primary calibrators, 3C48, 3C286, and~NGC7027~\cite{Baars1977, Ott1994} which are included in the~observation.

It takes about 5 min for each source with 4 + 4 observing sub-scans, and~about 2 h for nearly 25 sources to complete one cycle of observation, if~the sources are up (>$10^{\circ}$ elevation) in the sky, and~then we repeat the cycle of observation. So the sampling rate is about 0.5 per hour for each~source.

To analyze the variability, we used the statistical quantities, such as the modulation index $m$ (a measure of the relative strength of variations), {the modified modulation index or fractional variability~\cite{Schleicher2019} $m'$ (a parameter that similar to $m$ but take into account the error of flux density, when the part within the square root is less than 0, we set $m' = 0$, which imply that there is no variability), the~variability amplitude $Y$ which is assumed to be unbiased to the systematic noise}, and~the reduced chi-squared $\chi_{r}^2$, to~describe the variability as shown below, also see~\cite{Kraus2003},
\begin{equation}\label{fc:m}
m[\%] = 100\frac{\sigma_{s}}{<S>},
\end{equation}
\begin{equation}\label{fc:m'}
m'[\%] = 100\sqrt{\frac{\sigma_{s}^{2} - <\Delta S_{i}^{2}>}{<S> ^{2}}},
\end{equation}
\begin{equation}\label{fc:Y'}
Y[\%] = 3\sqrt{(m)^2-(m_0)^2},
\end{equation}
\begin{equation}\label{fc:chi2}
\chi_r^2 = \frac{1}{N-1}\sum(\frac{S_{i}-<S>}{\Delta S_{i}})^2,
\end{equation}
where $\sigma_{s}$ is the standard deviation of flux densities from the mean flux density ($<S>$) in a light curve, $m_{0}$ is the mean modulation index of all calibrators in the same observation, $N$ is the number of measurements in a light curve, $S_{i}$ and $\Delta S_{i}$ denote individual flux density and its~error.

\section{Results and~Discussion}
\unskip
\subsection{Variability~Characteristics}

\begin{figure}[H]
\widefigure
\includegraphics[width=8cm]{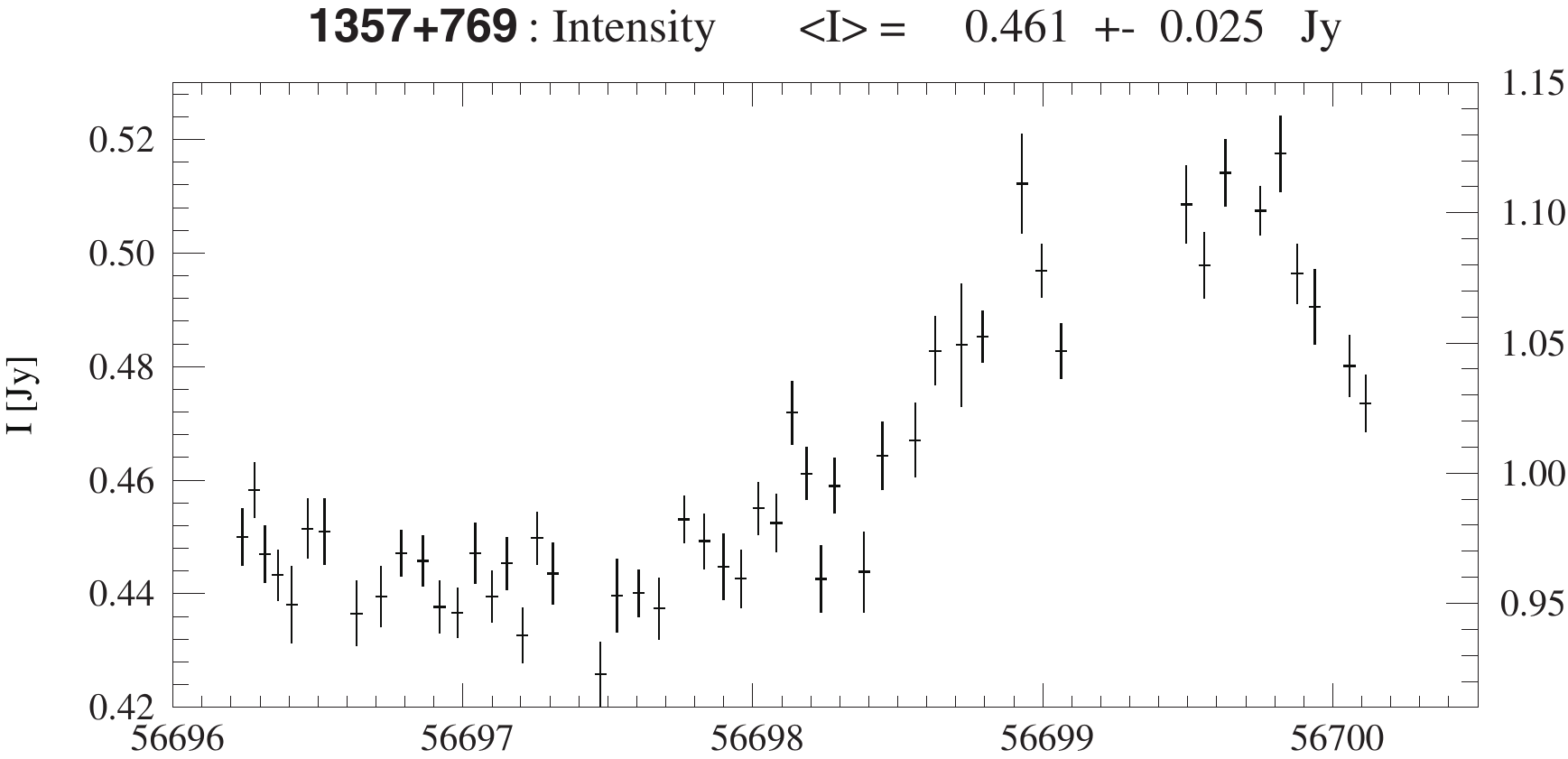}
\includegraphics[width=8cm]{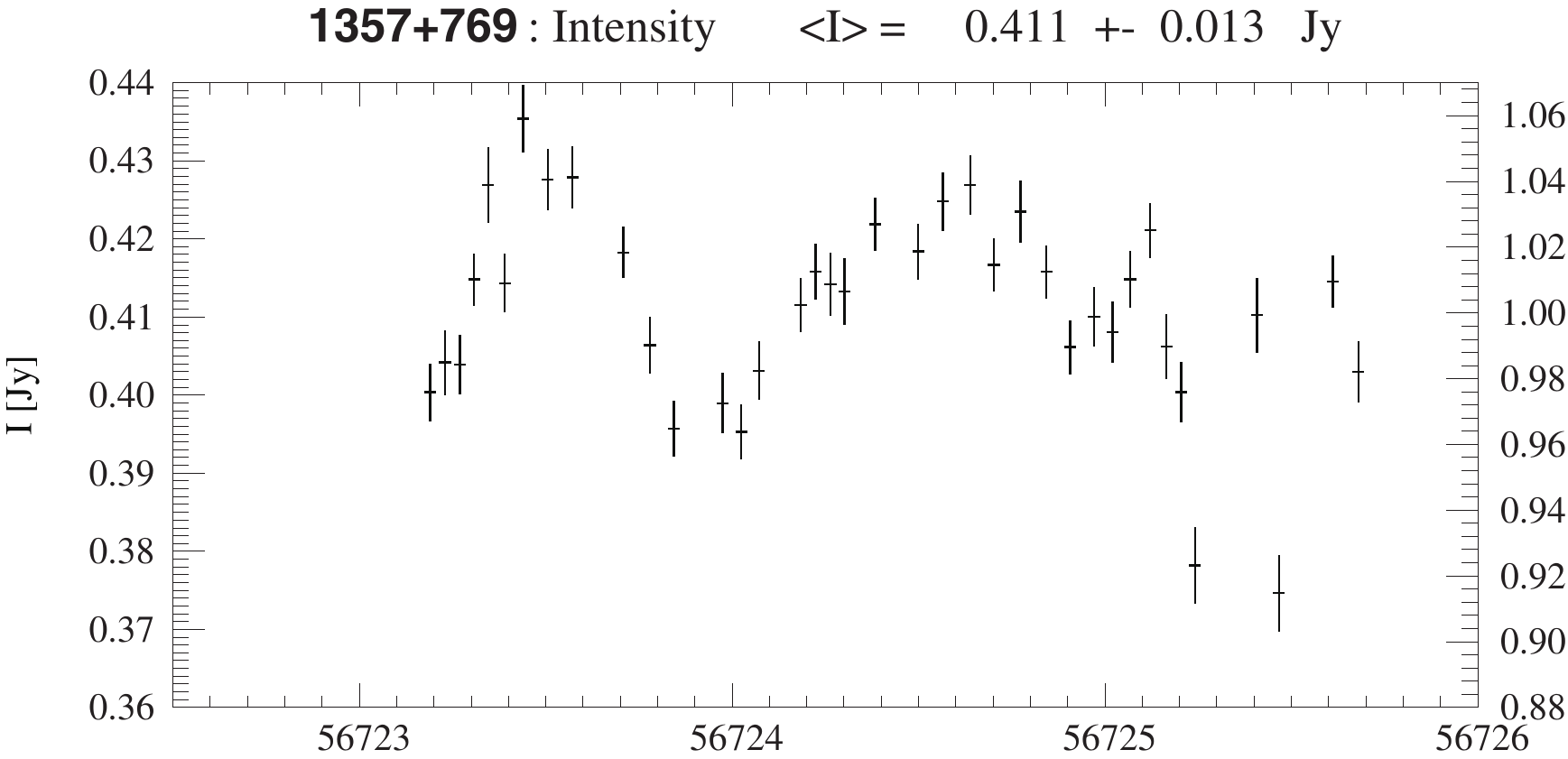}
\includegraphics[width=8cm]{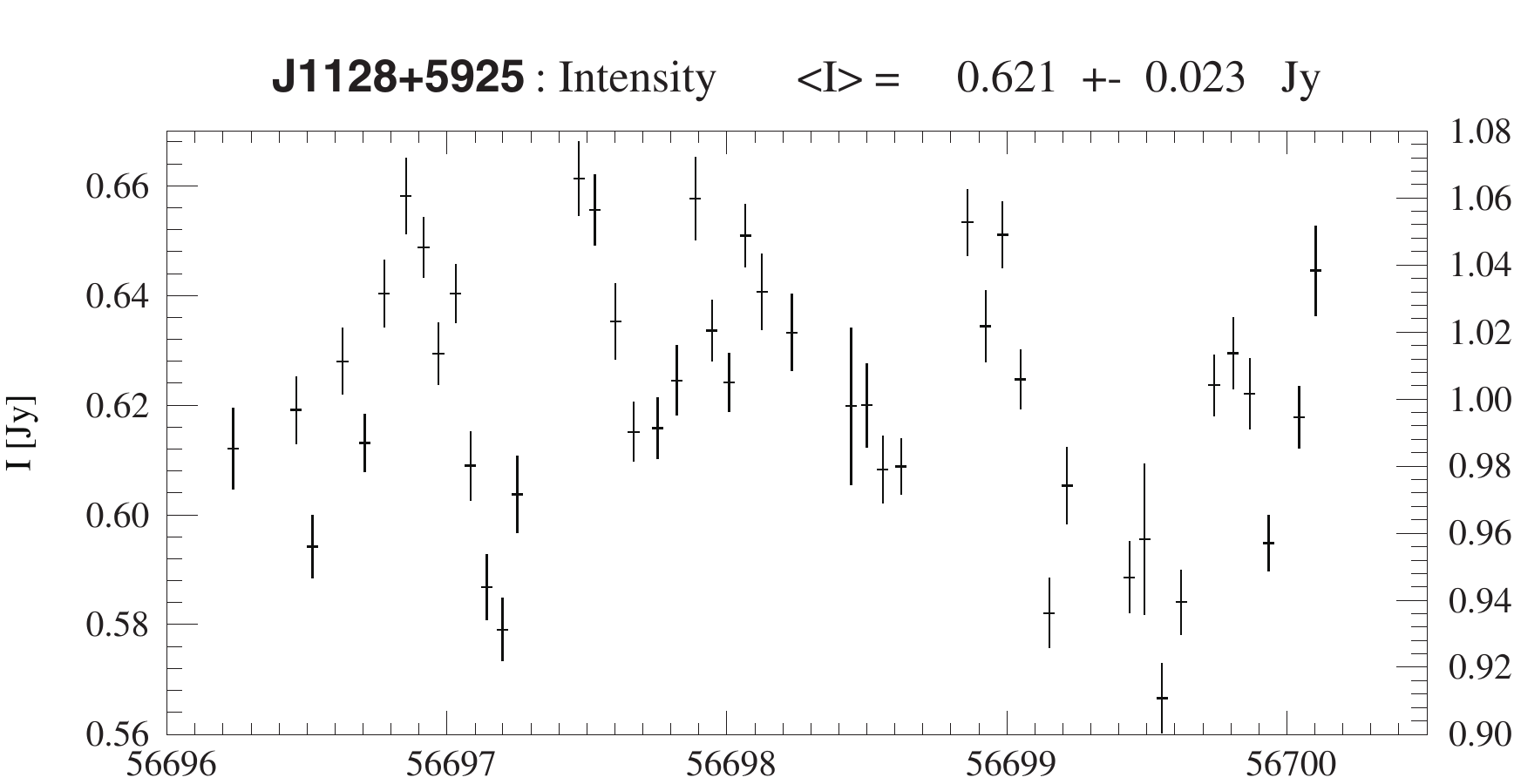}
\includegraphics[width=8cm]{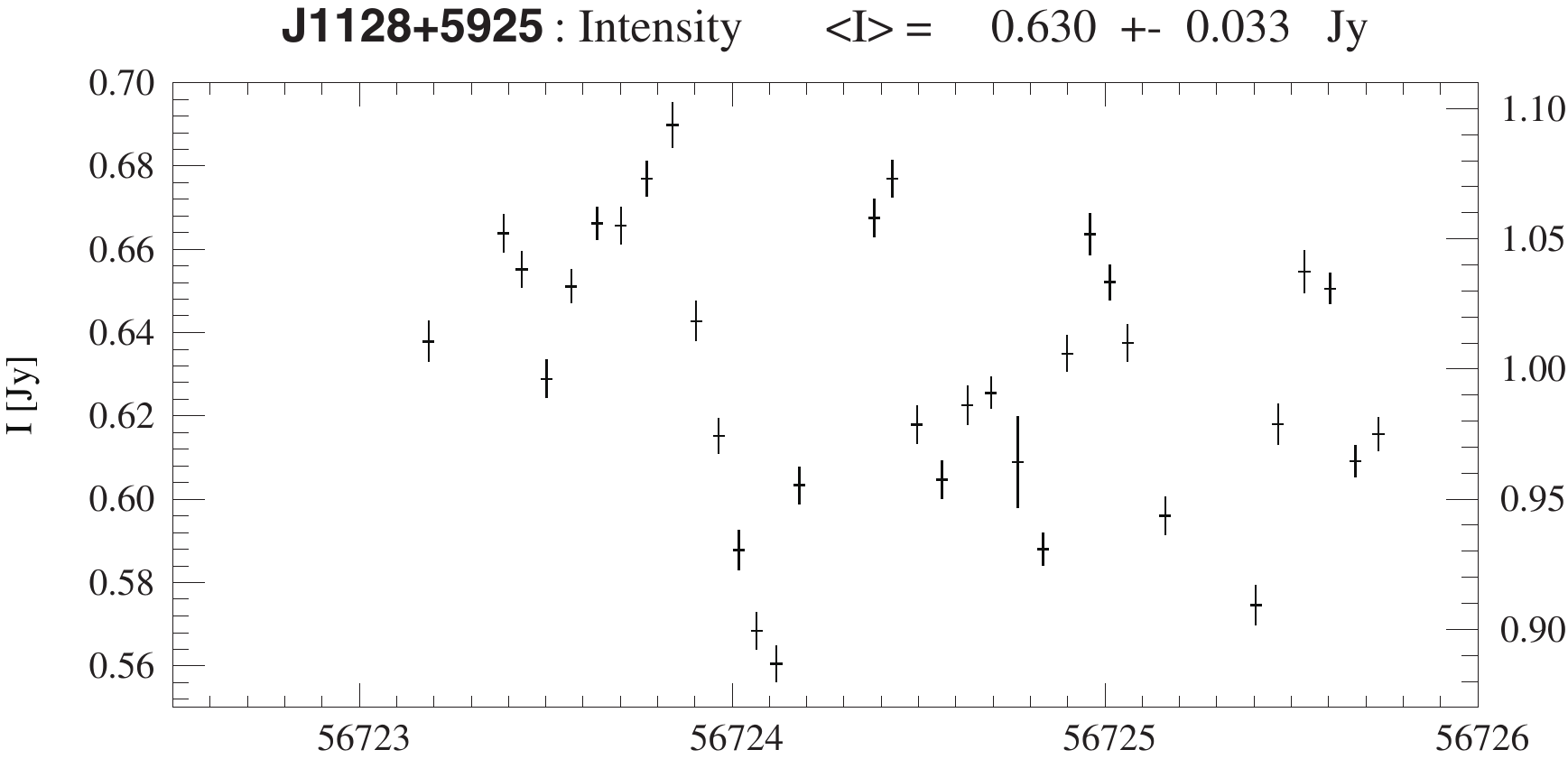}
\includegraphics[width=8cm]{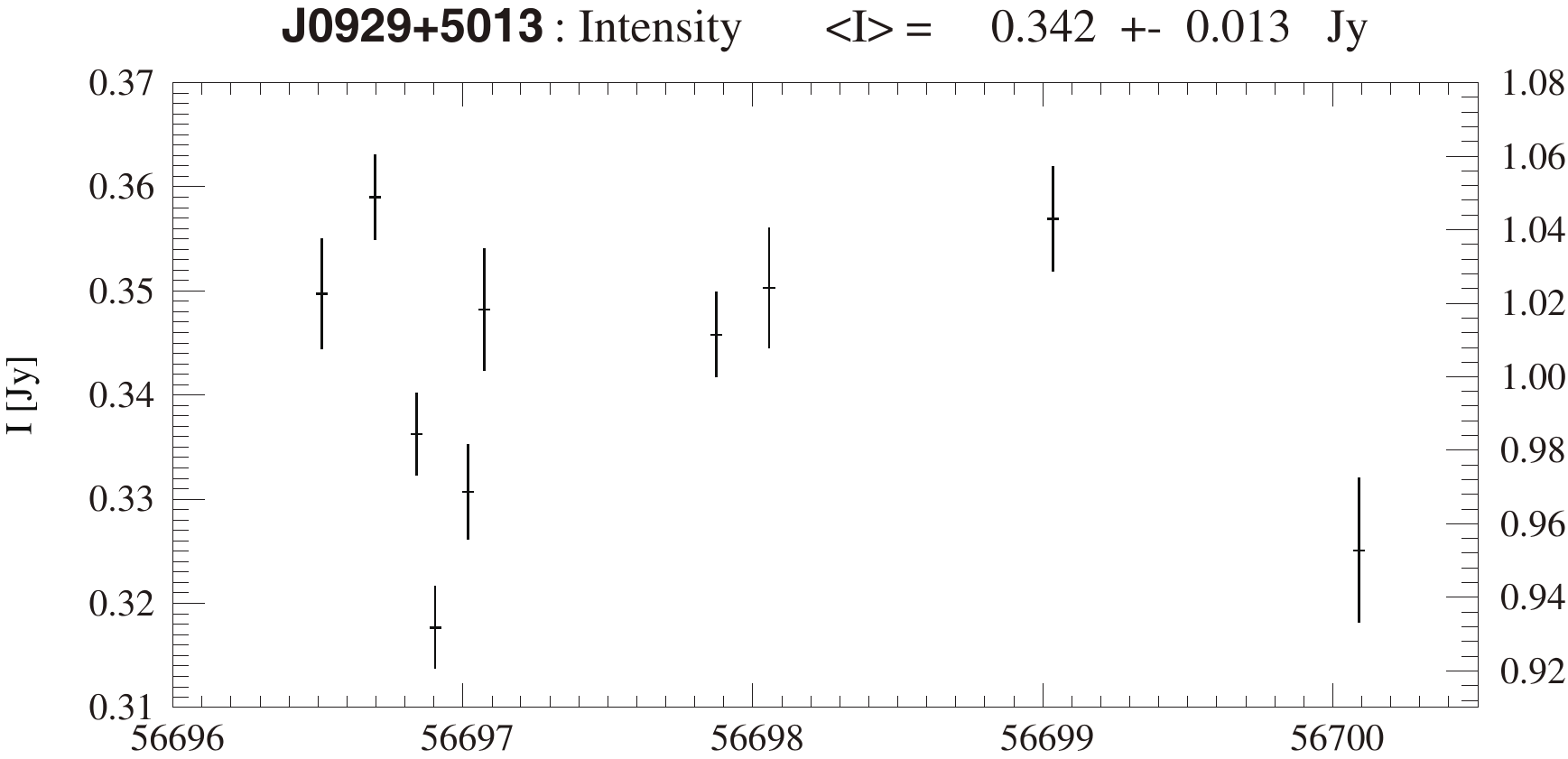}
\includegraphics[width=8cm]{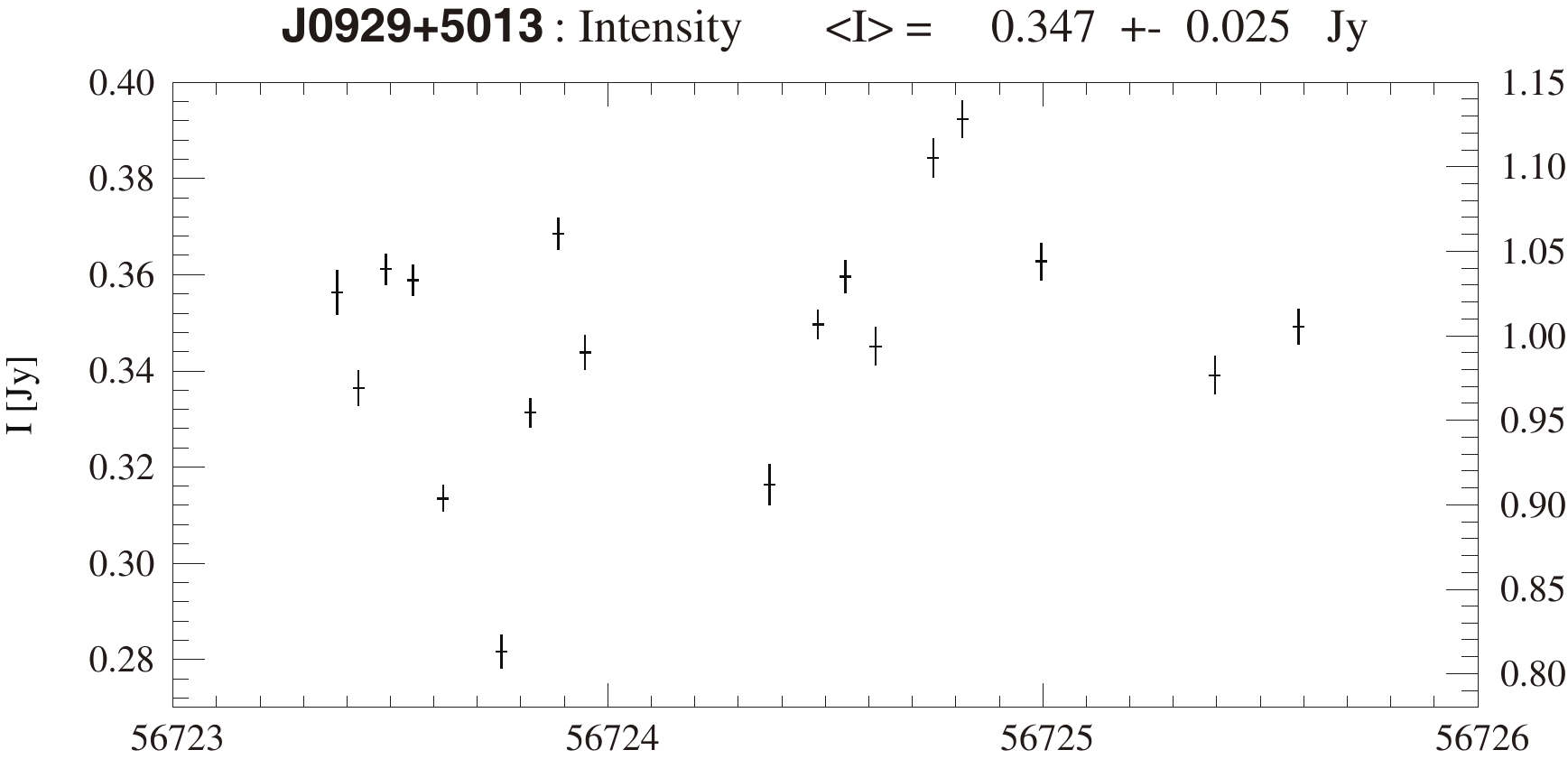}
\caption{{Three strong} 
IDV sources with $m \ge 3$ in both observations. ({\bf Left upper panel}) the light curve of 1357 + 769 at \mbox{4.8 GHz} in 8--12 February 2014, the~flux density $I$ (left $y$-axis) and the ratio of flux density over mean flux density $<I>$ (right $y$-axis) versus time (MJD) in this and following figures; ({\bf right upper panel}) the light curve of 1357 + 769 at 4.8 GHz in \mbox{7--9 March 2014}. ({\bf Left middle panel}) the light curve of J1128 + 5925 at 4.8 GHz in 8--12 February 2014; ({\bf right middle panel}) the light curve of J1128 + 5925 at 4.8 GHz in 7--9 March 2014.  ({\bf Left bottom panel}) the light curve of J0929 + 5013 at 4.8 GHz in 8--12 February 2014; ({\bf right bottom panel}) the light curve of J0929 + 5013 at 4.8 GHz in 7--9 March~2014.}
\label{fig1}
\end{figure}

\begin{figure}[H]
\widefigure
\includegraphics[width=8cm]{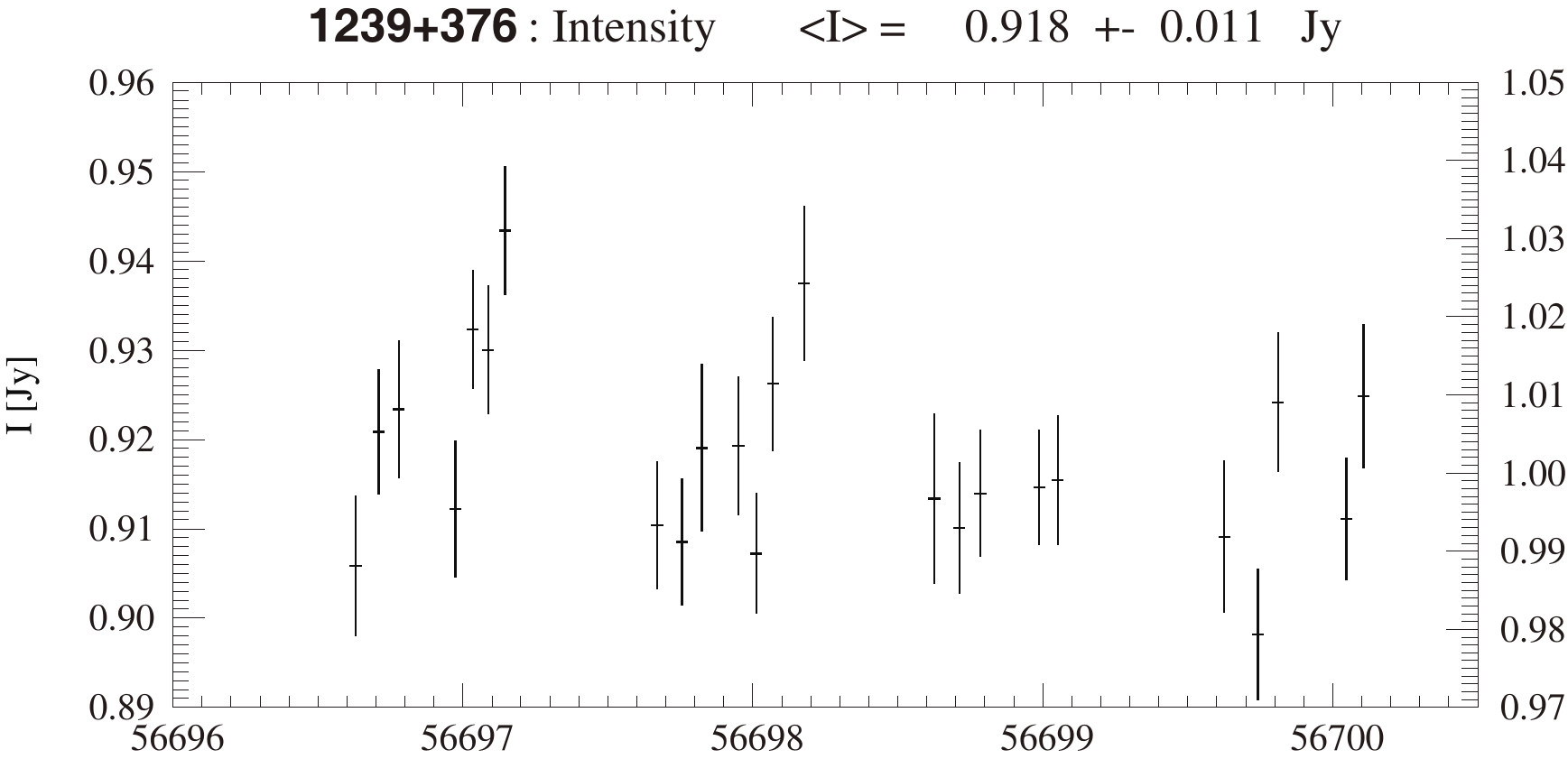}
\includegraphics[width=8cm]{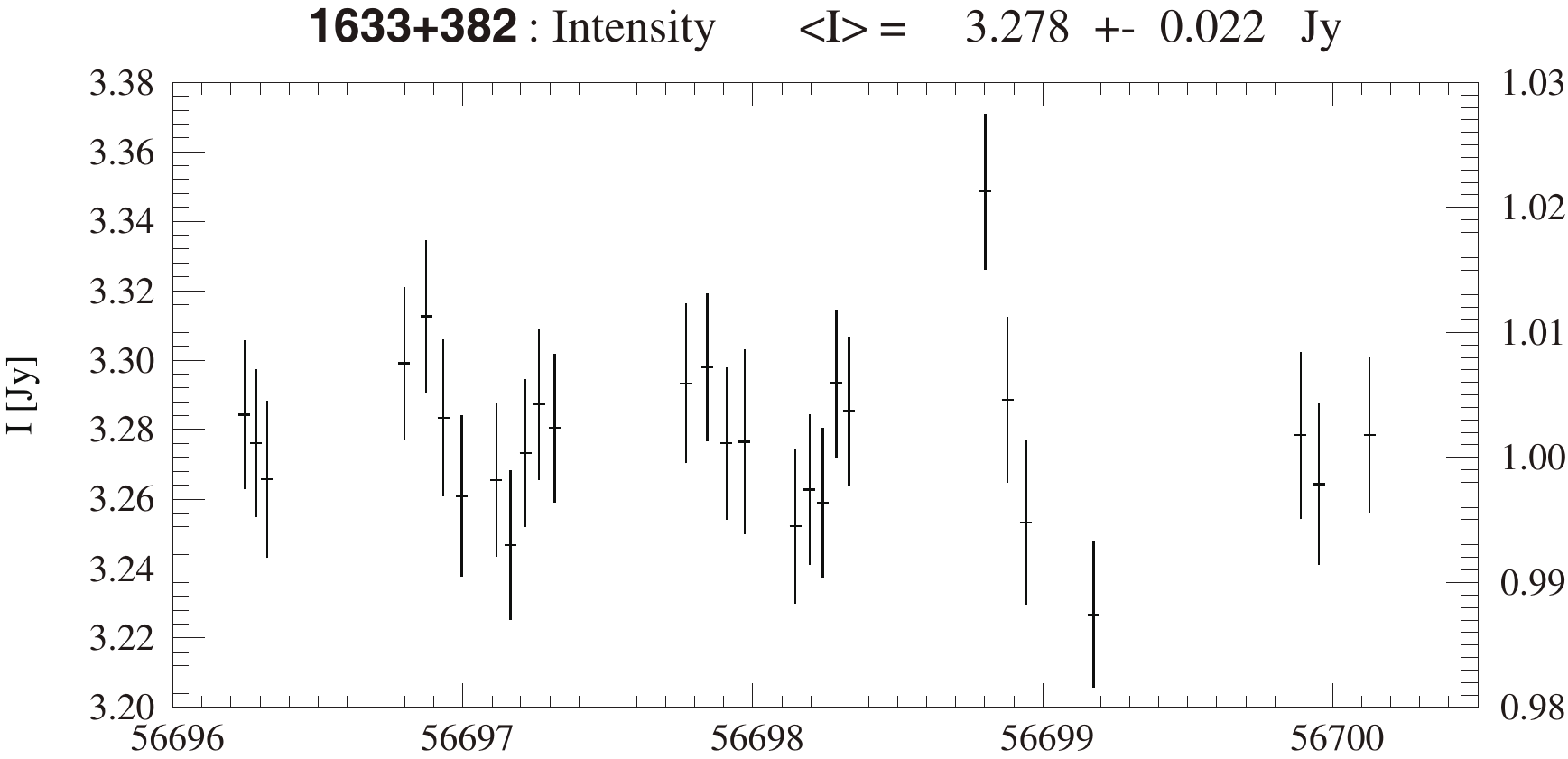}
\caption{{Two example} 
sources with no variability. ({\bf Left panel}) the light curve of B2 1239 + 376 at 4.8 GHz in  {8--12 February 2014}; ({\bf right panel}) the light curve of 1633 + 382 at 4.8 GHz in 8--12 February~2014.}
\label{fig2}
\end{figure}

The results are listed in Tables~\ref{tab1} and~\ref{tab2}. In~the February IDV session, 16 targets from the RA observing session are observed, plus 4 additional targets from other samples and 5 calibrators. In~the March IDV session, 14 targets from the RA observing session are observed, plus 3 additional targets from other samples and 5 calibrators. There are more than 20 data points sampled for the majority of sources in about three-day observation of each session, some sources have fewer data points due to their low declinations or bad data being deleted by the calibration software (e.g., when Full Width of Half Power of source profile > 120\% of antenna beam width, or~antenna pointing error > 1/6 of beam width, or~error of Gaussian fit > 10\% of peak of the Gaussian, or~amplitude of a sub-scan profile > 110\% of average amplitude of 4 + 4 sub-scans; these can be caused by too large source, too weak source or interference, etc.). After~careful calibrations as introduced above, we obtained flux densities and light curves for each source. {Here, we only show a few sources for example,} three sources with strong variability---1357 + 769, J1128 + 5925 and WISE J092915.43 + 501336.0 in Figure~\ref{fig1}, two counterexamples---B2 1239 + 376 and 1633 + 382 with no sign of variability in Figure~\ref{fig2}.

The same as our previous work~\cite{Liux2012b}, in~order to ascertain whether or not an IDV is present, we applied a $\chi^2$ test to each data set. We adopt the criterium that a data set with `a probability of $\leq$0.1\% to be constant' is considered to be variable (i.e., the `null hypothesis' probability $<$1.0 $\times\,10^{-3}$ in Table~\ref{tab1} or Table~\ref{tab2}), which is equivalent to the confidence level of 99.9\% to define the IDV. According to this definition, there are 14 out of 20 targets show IDV in session 1, all 17 targets show IDV in session~2.


In Tables~\ref{tab1} and~\ref{tab2}, we find there are a couple of prominent IDV sources from the two observing sessions, e.g.,~those with the modulation index $m>3 m_{0}$, the~$m_{0}$ is the average modulation index of the calibrators, which is 0.64 for session 1, and~0.45 for session~2.

The modified modulation index $m'$ represents a variability strength, but~not a real variability amplitude. The~parameter $Y$ is more like a variability amplitude, which ranges from a few per cent to $\sim$22\% for the prominent IDV sources in Tables~\ref{tab1} and~\ref{tab2}.

We note that there are 11 common targets appeared in both session 1 and session 2, all of them have shown IDV in the two sessions except the quasar 1156 + 295 which only shows IDV in session 2, we give a short discussion for this in Section~\ref{sec:discuss}.

Some of the IDV sources show relatively fast variability, e.g.,~the source J1128 + 5925, WISE J092915.43 + 501336.0 exhibit a short timescale of a few hours and some of them e.g.,~the source 1357 + 769 shows a relatively longer timescale. A~full analysis on the IDV timescales along with the source structure and brightness temperature derived from the RA space VLBI measurements will be presented in a forthcoming~paper.

\clearpage
\end{paracol}
\nointerlineskip
\unskip

\begin{paracol}{2}
\switchcolumn
\vspace{-6pt}

\subsection{Comment on Individual~Sources} \label{sec:sources}

We comment briefly on some prominent IDV sources which showed a modulation index $m>3m_{0}$ in one or two observing sessions, in~the following. We note that all the target sources are blazars with core-dominated radio structures in milliarcsecond scales from the VLBI images (e.g., \cite{Lister2016, Gabanyi2009}).

{\bf PKS 0528 + 134}: the source shows IDV in both sessions, see Tables~\ref{tab1} and~\ref{tab2}, the~IDV strength is slightly increased from session 1 to session 2, while the mean flux density of 1.710 $\pm$ 0.034 Jy in session 1 decreases to 1.525 $\pm$ 0.034 Jy, at~4.8~GHz.

{\bf 0716 + 714}: this BL Lac object is a well known IDV source, although~its IDV strength is moderate with relatively longer IDV timescales from previous observations and its IDV timescales can be fitted with the annual modulated ISS model~\cite{Liux2012b}. It shows IDV in both sessions, and~its IDV strength in session 1 is less than that in session 2. The~total flux density of 1.255 $\pm$ 0.016 Jy in session 1 decreases to 1.193 $\pm$ 0.017 Jy in session~2.

{\bf 0748 + 126}: this blazar shows IDV in both sessions, its IDV strength in session 1 is higher than that in session 2. Its mean flux density is similar from session 1 to session~2.

{\bf OJ287}: this is a BL Lac object, which is thought to be a supermassive binary black hole system from long-term optical periodic flux variations~\cite{Valtonen2007}, but~even the RA space VLBI still cannot resolve the binary. It is observed in session 2 only and shows an~IDV.

{\bf 1044 + 719}: this blazar shows IDV in both sessions. The~IDV strength increases from session 1 to session 2. Its mean flux density also increases from 2.581 $\pm$ 0.035 Jy in session 1 to 2.786 $\pm$ 0.049 Jy in session~2.



{\bf 1357 + 769}: this blazar shows strong IDVs in both sessions (upper panel of Figure~\ref{fig1}), the~$m'$ decreases from 5.28 in session 1 to 3.02 in session 2, while the mean flux density decreases from 0.461 $\pm$ 0.025 Jy in session 1 to 0.411 $\pm$ 0.013 Jy in session~2.

\startlandscape
\begin{specialtable}[H]
\widetable
\caption[]{The result of the IDV observation in 8--12 February 2014 with Urumqi 25 m telescope at 4.8~GHz. 		The columns are: (1) name; (2) source type (FSRQ: flat spectrum radio quasar, BL: BL Lac object, CSSQ: compact steep spectrum quasar, G: galaxy, PN: planetary nebula), most target sources' type can be found in~\cite{Healey2007}, except~three targets with additional references in footnote; (3) number of flux density measurements; (4) mean flux density with the standard deviation; (5) modulation index $m$ and modified modulation index $m'$; (6) relative variability amplitude $Y$; (7), (8) $\chi^2$ and reduced $\chi^2$; (9) the null probability (a probability that the light curve can be fitted with a constant value) and with detected IDV (y) or not (n) in the bracket; (10) redshift of source; (11) previously detected IDV (y) or not (n) with relative reference in the bracket, or~no available observation/publication found (n.a.). The~sources in the first part are from RadioAstron observing list, the~sources in the second part are from other samples, the~sources in the third part are calibrators which did not show IDV from previous observations.}
\setlength{\tabcolsep}{6.8mm}
\scalebox{.85}[.85]{
\begin{threeparttable}
\begin{tabular}{ccccccccccc}
\toprule
\noalign{\smallskip}
\textbf{1} & \textbf{2}	& \textbf{3} & \textbf{4} & \textbf{5} & \textbf{6} & \textbf{7} & \textbf{8} & \textbf{9} & \textbf{10} & \textbf{11} 	\\

\textbf{Source}	& \textbf{Type}	& \textbf{N}		& \bm{$<S> \pm\,\sigma_{s}$}	& \bm{$m/m'$}	& \bm{$Y$}	& \bm{$\chi^2$}	& \bm{$\chi_r^2$}	& \bm{$P_{null}$} \bf(y/n)	& \textbf{Redshift}	& \textbf{y/n (Ref.)}	\\

&	& 	& \textbf{[Jy]}	& \textbf{[\%]}	& \textbf{[\%]}	&	&	&	&   &	\\

\midrule
\noalign{\smallskip}

0202 + 319				& FSRQ			& 27 	& 1.675 $\pm$ 0.020  & 1.19/0.95 	& 3.02  & 74.8  & 2.9  & {$1.3\times10^{-06}$}  (y) * 
& 1.466		& n.a. 	\\
0300 + 470				& $\rm BL\,^{a}$& 21 	& 2.548 $\pm$ 0.029  & 1.14/0.91 	& 2.81  & 60.1  & 3.0  & $6.8\times10^{-06}$ (y) *& 0.475		& nnnn (1) 	\\
B2 0340 + 362			& FSRQ			& 10 	& 0.452 $\pm$ 0.005  & 1.11/0.00 	& 2.70  & 9.5   & 1.1  & $3.9\times10^{-01}$ (n) & 1.484		& yyyy (1) 	\\
PKS 0507 + 179			& FSRQ			& 15 	& 0.584 $\pm$ 0.008  & 1.37/0.94 	& 3.63  & 30.6  & 2.2  & $6.4\times10^{-03}$ (n) & 0.41		& yyyn (1) 	\\
PKS 0528 + 134			& FSRQ			& 22 	& 1.710 $\pm$ 0.034  & 1.99/1.83 	& 5.64  & 149.8 & 7.1  & $1.9\times10^{-21}$ (y) & 2.06		& nyny (1) 	\\
0642 + 449				& FSRQ			& 23 	& 2.553 $\pm$ 0.027  & 1.06/0.79	& 2.52  & 55.6  & 2.5  & $9.7\times10^{-05}$ (y) *& 3.396		& nnnn (1) 	\\
0716 + 714				& BL			& 43 	& 1.255 $\pm$ 0.016  & 1.27/0.99 	& 3.30  & 112.6 & 2.7  & $2.3\times10^{-08}$ (y) & 0.3		& y (2) 	\\
0748 + 126				& FSRQ			& 24 	& 2.559 $\pm$ 0.057  & 2.23/2.11 	& 6.40  & 237.9 & 10.3 & $1.3\times10^{-37}$ (y) & 0.889		& nyyy (1) 	\\
0917 + 624				& FSRQ			& 44 	& 1.123 $\pm$ 0.018  & 1.60/1.37 	& 4.40  & 186.0 & 4.3  & $1.1\times10^{-19}$ (y) & 1.446		& nnnn (1)	\\
0954 + 658				& FSRQ			& 43 	& 1.113 $\pm$ 0.017  & 1.53/1.29 	& 4.15  & 164.3 & 3.9  & $2.2\times10^{-16}$ (y) & 0.368		& nyyy (1) 	\\
1044 + 719				& FSRQ			& 15 	& 2.581 $\pm$ 0.035  & 1.36/1.15 	& 3.58  & 57.5  & 4.1  & $3.3\times10^{-07}$ (y) & 1.15		& nnyn (1) 	\\
1156 + 295				& FSRQ			& 26 	& 1.248 $\pm$ 0.013  & 1.04/0.70 	& 2.45  & 45.4  & 1.8  & $7.5\times10^{-03}$ (n) & 0.72469	& yyyy (1) 	\\
B2 1239 + 376			& FSRQ			& 24 	& 0.918 $\pm$ 0.011  & 1.20/0.00 	& 3.03  & 47.8  & 2.1  & $1.8\times10^{-03}$ (n) & 3.81908	& n.a	\\
1357 + 769				& FSRQ			& 51 	& 0.461 $\pm$ 0.025  & 5.42/5.28 	& 16.15 & 972.4 & 19.4 & $3.6\times10^{-17}$ (y) *& 1.585		& n.a.	\\
PKS 1502 + 106			& FSRQ			& 9		& 0.791 $\pm$ 0.005  & 0.63/0.00	& 0.00  & 4.1   & 0.5  & $8.5\times10^{-01}$ (n) & 1.83831	& n (3) 	\\
1637 + 574				& FSRQ			& 39	& 1.308 $\pm$ 0.016  & 1.22/0.93	& 3.12  & 86.0  & 2.3  & $1.4\times10^{-05}$ (y) *& 0.7506	& nnnn (1) 	\\

\midrule
\noalign{\smallskip}
WISE J092915.43 + 501336.0& $\rm BL\,^{b}$	& 10 	& 0.342 $\pm$ 0.013  & 3.80/3.50 	& 11.24 & 83.3  & 9.3  & $3.6\times10^{-14}$ (y) & 0.37039	& yyyy (1)	\\
BZQ J1128 + 5925		& FSRQ			& 46 	& 0.621 $\pm$ 0.023  & 3.70/3.54  & 10.94 & 616.5 & 13.7 & $2.0\times10^{-101}$ (y)& 1.795		& y (4) 	\\
1633 + 382				& FSRQ			& 28 	& 3.278 $\pm$ 0.022  & 0.67/0.00	& 0.56 	& 29.1  & 1.1  & $3.6\times10^{-01}$ (n) & 1.81309	& nyyn (1) 	\\
J2311 + 4543			& $\rm FSRS\,^{c}$& 24 	& 0.662 $\pm$ 0.016  & 2.42/2.22 	& 6.99  & 139.8 & 6.1  & $9.9\times10^{-19}$ (y) & 1.447		& yyyy (1) 	\\

\midrule
\noalign{\smallskip}
3C48					& CSSQ			& 33 	& 5.590 $\pm$ 0.038  & 0.68/0.06  	& 0.65 	& 35.3  & 1.1  & $3.1\times10^{-02}$ 	& 0.367		&  		\\
0836 + 710				& FSRQ			& 52 	& 3.434 $\pm$ 0.020  & 0.58/0.00	& 0.00  & 34.7  & 0.7  & $9.6\times10^{-01}$ 	& 2.172		& 	 	\\
0951 + 699 (M82)		& G				& 49 	& 3.557 $\pm$ 0.021  & 0.59/0.00	& 0.00  & 34.0  & 0.7  & $9.4\times10^{-01}$ 	& 0.00068	&  		\\
3C286					& CSSQ			& 28 	& 7.511 $\pm$ 0.046  & 0.61/0.00	& 0.00  & 22.9  & 0.8  & $6.9\times10^{-01}$ 	& 0.84993	&  		\\
NGC7027					& PN			& 26 	& 5.411 $\pm$ 0.041  & 0.76/0.37	& 1.20  & 34.5  & 1.4  & {$9.7\times10^{-02}$} 	& 0			&  		\\

\noalign{\smallskip}
\bottomrule
\end{tabular}{}
\begin{tablenotes}
\item {\it Notes.} References for sources' type: a.~\cite{VeronCetty2006}; b.~\cite{Plotkin2008}; c.~\cite{Lazio2008}. References for previously detected IDV or not (Column 11). 1.~\cite{Lovell2008}, where the classification per epoch of variable  (y) or nonvariable  (n) in four epochs (with each of 3 days) observed with the VLA at 5 GHz around year 2002; 2.~\cite{Liux2012a} 3.~\cite{Quirrenbach1992} 4.~\cite{Gabanyi2007}. The~* marked in Column 9 indicates new IDVs found from this observation.
\end{tablenotes}

\end{threeparttable}
}
\label{tab1}

\end{specialtable}

\begin{specialtable}[H]
\widetable
\caption{{The result} 
of the IDV observation in 7--9 March 2014 with Urumqi 25~m telescope at 4.8 GHz. The~columns are same as that in Table~\ref{tab1}. The~* marked in Column 9 indicates new IDVs found from this observation.}
\setlength{\tabcolsep}{6.9mm}
\begin{threeparttable}
\scalebox{.85}[.85]{
\begin{tabular}{ccccccccccc}
\toprule
\noalign{\smallskip}

\textbf{Source} & \textbf{Type} & \textbf{N}  & \bm{$<S> \pm \sigma_{s}$}  & \bm{$m/m'$}  & \bm{$Y$} & \bm{$\chi^2$} & \bm{$\chi_r^2$} & \bm{$P_{null}$} \bf{(y/n)} & \textbf{Redshift} & \textbf{y/n (Ref.)}\\

&  &  & \textbf{[Jy]} & \textbf{[\%]} & \textbf{[\%]}&  &  &  &  &\\

\midrule
\noalign{\smallskip}

PKS 0528 + 134				& FSRQ	& 23	& 1.525 $\pm$ 0.034	& 2.23/2.17	& 6.55	& 405.8	& 18.4	& {$2.6\times10^{-72}$ (y)}
& 2.06		& nyny (1)	\\
4C  + 72.10 (0604 + 728)	& FSRQ	& 43	& 0.622 $\pm$ 0.009	& 1.45/1.23	& 4.12	& 158.9	& 3.8	& $1.7\times10^{-15}$ (y) *	& 3.53		& n.a.	\\
0716 + 714					& BL	& 48	& 1.193 $\pm$ 0.017	& 1.42/1.31	& 4.05	& 322.0	& 6.9	& $1.2\times10^{-42}$ (y)	& 0.3		& y (2)	\\
0748 + 126					& FSRQ	& 19	& 2.525 $\pm$ 0.043	& 1.70/1.63	& 4.93	& 224.0	& 12.4	& $1.5\times10^{-37}$ (y)	& 0.889		& nyyy (1)	\\
OJ287						& BL	& 19	& 2.881 $\pm$ 0.053	& 1.84/1.77	& 5.35	& 274.0	& 15.2	& $1.1\times10^{-47}$ (y)	& 0.3056	& y (3)	\\
0917 + 624					& FSRQ	& 48	& 1.112 $\pm$ 0.014	& 1.26/1.10	& 3.53	& 206.4	& 4.4	& $7.4\times10^{-22}$ (y)	& 1.446		& nnnn (1)	\\
0953 + 254					& FSRQ	& 20	& 1.149 $\pm$ 0.013	& 1.13/0.96	& 3.11	& 70.6	& 3.7	& $7.4\times10^{-08}$ (y)	& 0.70721	& nyyn (1)	\\
0954 + 658					& FSRQ	& 48	& 1.038 $\pm$ 0.013	& 1.25/1.09	& 3.50	& 230.2	& 4.9	& $5.6\times10^{-26}$ (y)	& 0.368		& yyyy (1)	\\
1044 + 719					& FSRQ	& 48	& 2.786 $\pm$ 0.049	& 1.76/1.68	& 5.10	& 599.7	& 12.8	& $6.6\times10^{-97}$ (y)	& 1.15		& nnyn (1)	\\
B2 1128 + 385				& FSRQ	& 21	& 1.385 $\pm$ 0.020	& 1.44/1.33	& 4.12	& 144.4	& 7.2	& $7.3\times10^{-21}$ (y)	& 1.74046	& nyny (1)	\\
1156 + 295					& FSRQ	& 19	& 1.236 $\pm$ 0.011	& 0.89/0.69	& 2.30	& 48.9	& 2.7	& $1.1\times10^{-04}$ (y)	& 0.72469	& yyyy (1)	\\
1357 + 769					& FSRQ	& 38	& 0.411 $\pm$ 0.013	& 3.16/3.02	& 9.39	& 339.1	& 9.2	& $1.8\times10^{-50}$ (y)	& 1.585		& n.a.	\\
1611 + 343					& FSRQ	& 21	& 3.956 $\pm$ 0.045	& 1.14/1.02	& 3.13	& 113.2	& 5.7	& $5.1\times10^{-15}$ (y)	& 1.39915	& nynn (1)	\\
3C345						& FSRQ	& 21	& 6.366 $\pm$ 0.057	& 0.90/0.75	& 2.32	& 75.4	& 3.8	& $2.4\times10^{-08}$ (y) *	& 0.5928	& nnnn (1)	\\

\midrule
WISE J092915.43 + 501336.0	& BL	& 18	& 0.347 $\pm$ 0.025	& 7.20/7.13	& 21.57	& 879.2	& 51.7	& $5.9\times10^{-176}$ (y)	& 0.37039	& yyyy (1)	\\
BZQ J1128 + 5925				& FSRQ	& 34	& 0.630 $\pm$ 0.033	& 5.24/5.18	& 15.66	& 1721.5& 52.2	& $0.0\times10^{-00}$ (y)	& 1.795		& y (4)	\\
J2311 + 4543					& FSRQ	& 21	& 0.671 $\pm$ 0.016	& 2.38/2.27	& 7.02	& 209.4	& 10.5	& $1.5\times10^{-33}$ (y)	& 1.447		& yyyy (1) 	\\

\midrule
3C48						&		& 27	& 5.604 $\pm$ 0.028	& 0.50/0.12	& 0.64	& 28.5	& 1.1	& $3.3\times10^{-01}$	&			&  		\\
0836 + 710					&		& 48	& 3.404 $\pm$ 0.014	& 0.41/0.00	& 0.00	& 32.9	& 0.7	& $9.4\times10^{-01}$	&			& 		\\
0951 + 699 (M82)			&		& 48	& 3.548 $\pm$ 0.014	& 0.39/0.00	& 0.00	& 28.3	& 0.6	& $9.9\times10^{-01}$	&			&  		\\
3C286						&		& 18	& 7.506 $\pm$ 0.027	& 0.36/0.00	& 0.00	& 8.8	& 0.5	& $9.5\times10^{-01}$	&			&  		\\
NGC7027						&		& 25	& 5.406 $\pm$ 0.032	& 0.59/0.00	& 1.15	& 37.2	& 1.6	& {$4.2\times10^{-02}$}	&			&  		\\
\noalign{\smallskip}
\bottomrule
\end{tabular}{}
}


\end{threeparttable}
\label{tab2}
\end{specialtable}

\finishlandscape


{\bf WISE J092915.43 + 501336.0 (J0929 + 5013, B0925 + 504)}: the BL Lac object exhibits fast IDV in both sessions (bottom panel of Figure~\ref{fig1}), $m'$ increases from 3.50 in session 1 to 7.13 in session 2. Its mean flux density is slightly increased from session 1 to \mbox{session 2}. The~blazar has previously shown fast IDV with Urumqi 25~m and Efferlsberg 100~m telescope, its IDV timescales can be fitted with the annual modulated ISS model~\cite{LiuLiu2015, Liux2017}.

{\bf BZQ J1128 + 5925}: the blazar shows a strong fast IDV in both sessions (middle panel of Figure~\ref{fig1}), its $m'$ increases from 3.54 in session 1 to 5.18 in session 2. Its mean flux density increases slightly from 0.621 $\pm$ 0.023 Jy in session 1 to 0.630 $\pm$ 0.033 Jy in session 2. This fast IDV was first found with Urumqi 25~m telescope, and~it has shown seasonal variations of IDV timescales and can be well fitted with the annual modulation model~\cite{Gabanyi2007}.

{\bf J2311 + 4543}: the blazar shows a strong IDV in both sessions with both the $m'$ and mean flux density slightly increased from session 1 to session~2.
\subsection{Discussion} \label{sec:discuss}



We have tried to cross-correlate the parameters listed in Tables~\ref{tab1} and~\ref{tab2}. No correlation is found between the IDV strength ($m'$) and either redshift or Galactic latitude, as~shown in Figure~\ref{fig3}a,b, with~correlation parameters listed in Table~\ref{tab3}. Three strong IDVs appear in area of relatively lower flux density $<S> < 0.7$ Jy as shown in Figure~\ref{fig4}.


\end{paracol}
\nointerlineskip
\begin{figure}[H]
\widefigure
\subfigure{\label{fig3a}
\includegraphics[width=8cm]{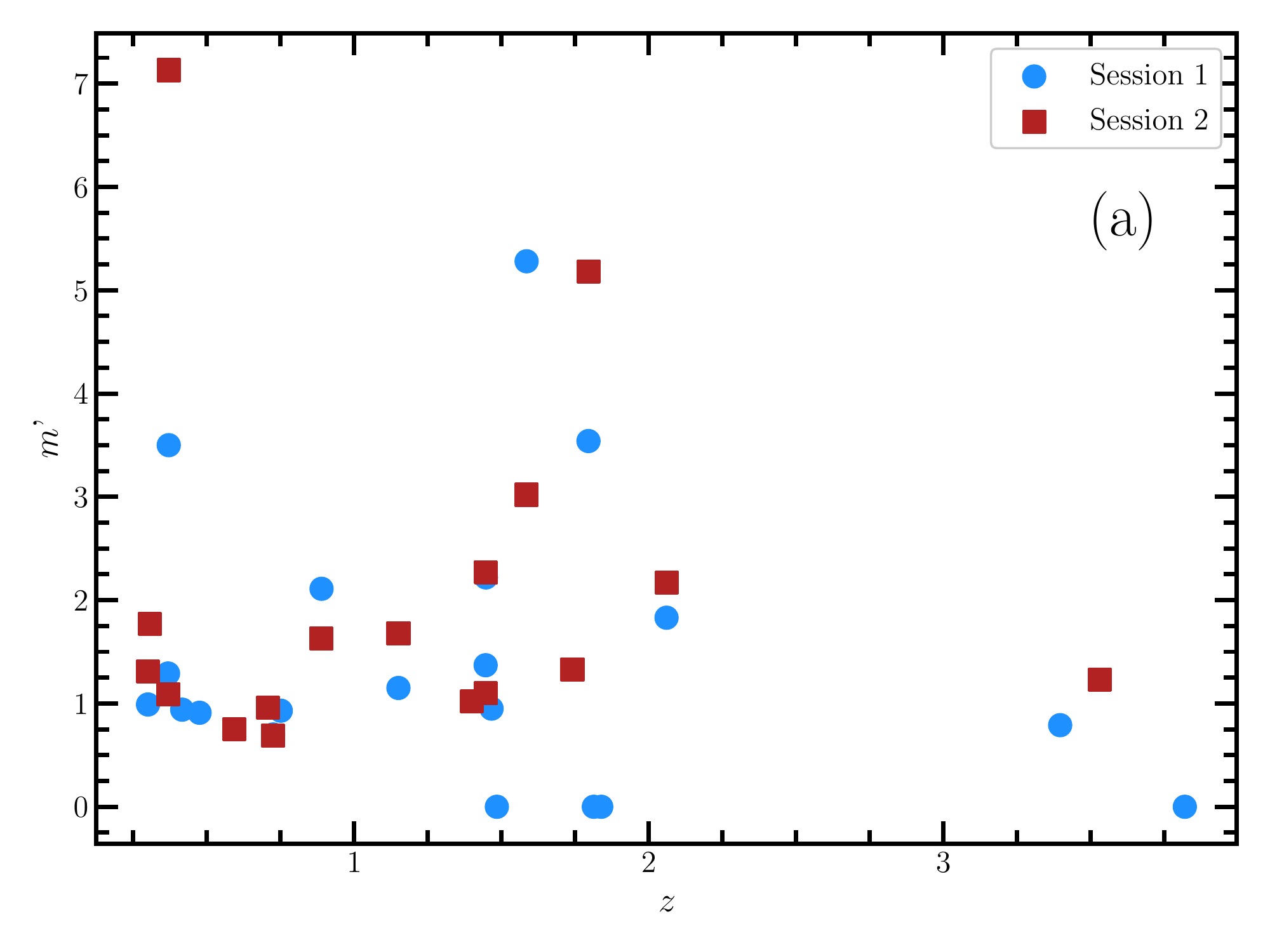}}
\subfigure{\label{fig3b}
\includegraphics[width=8cm]{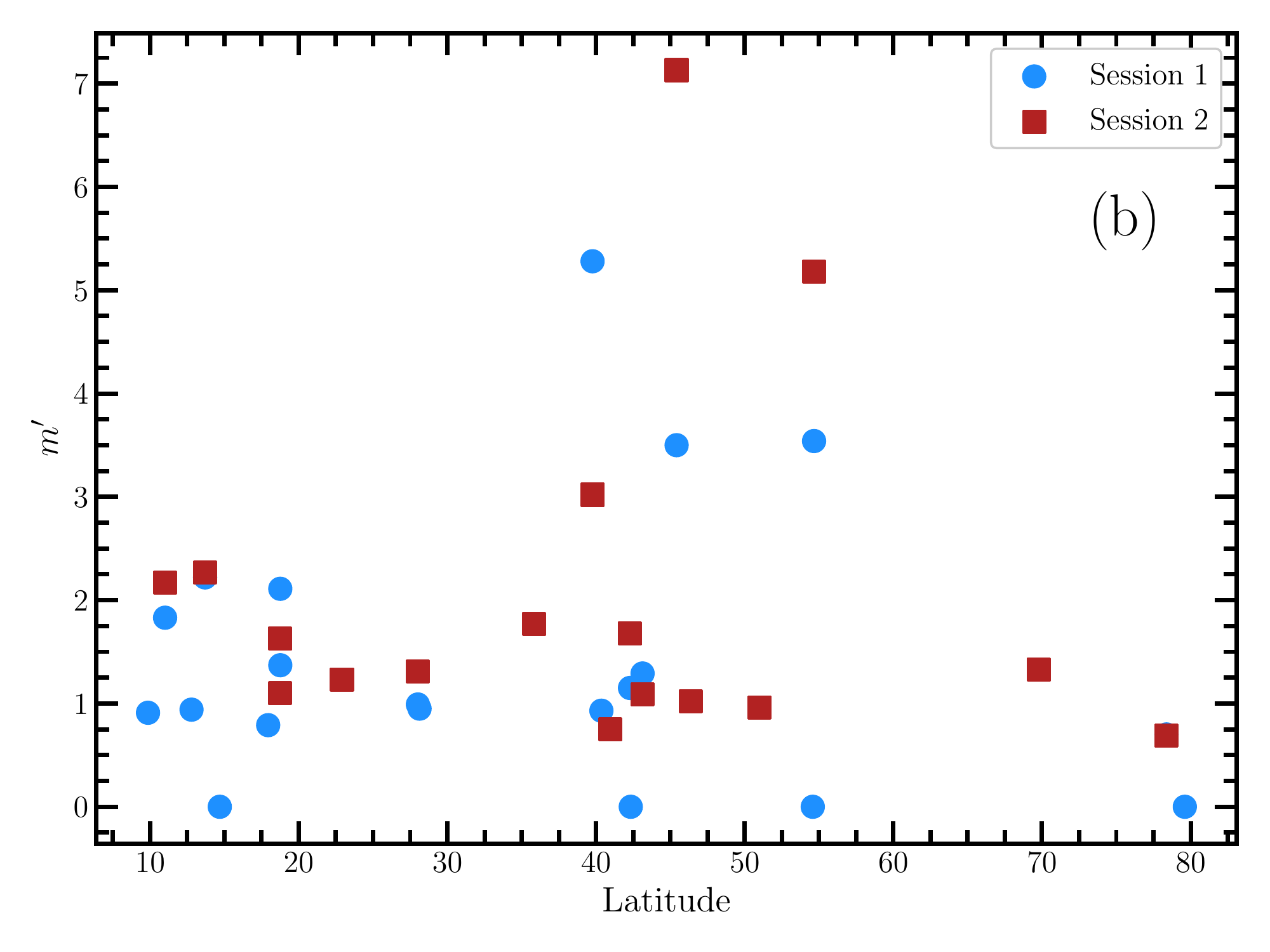}}
\caption{({\bf a}) {The modified} 
modulation index ($m'$ in per cent) versus redshift ($z$) of the target sources observed in session 1 (blue circle) and session 2 (red square); ({\bf b}) the modified modulation index ($m'$ in per cent) versus Galactic latitude of the target sources observed in session 1 (blue circle) and session 2 (red square).}
\label{fig3}
\end{figure}
\begin{paracol}{2}
\switchcolumn

\begin{figure}[H]
\includegraphics[width=7.5cm]{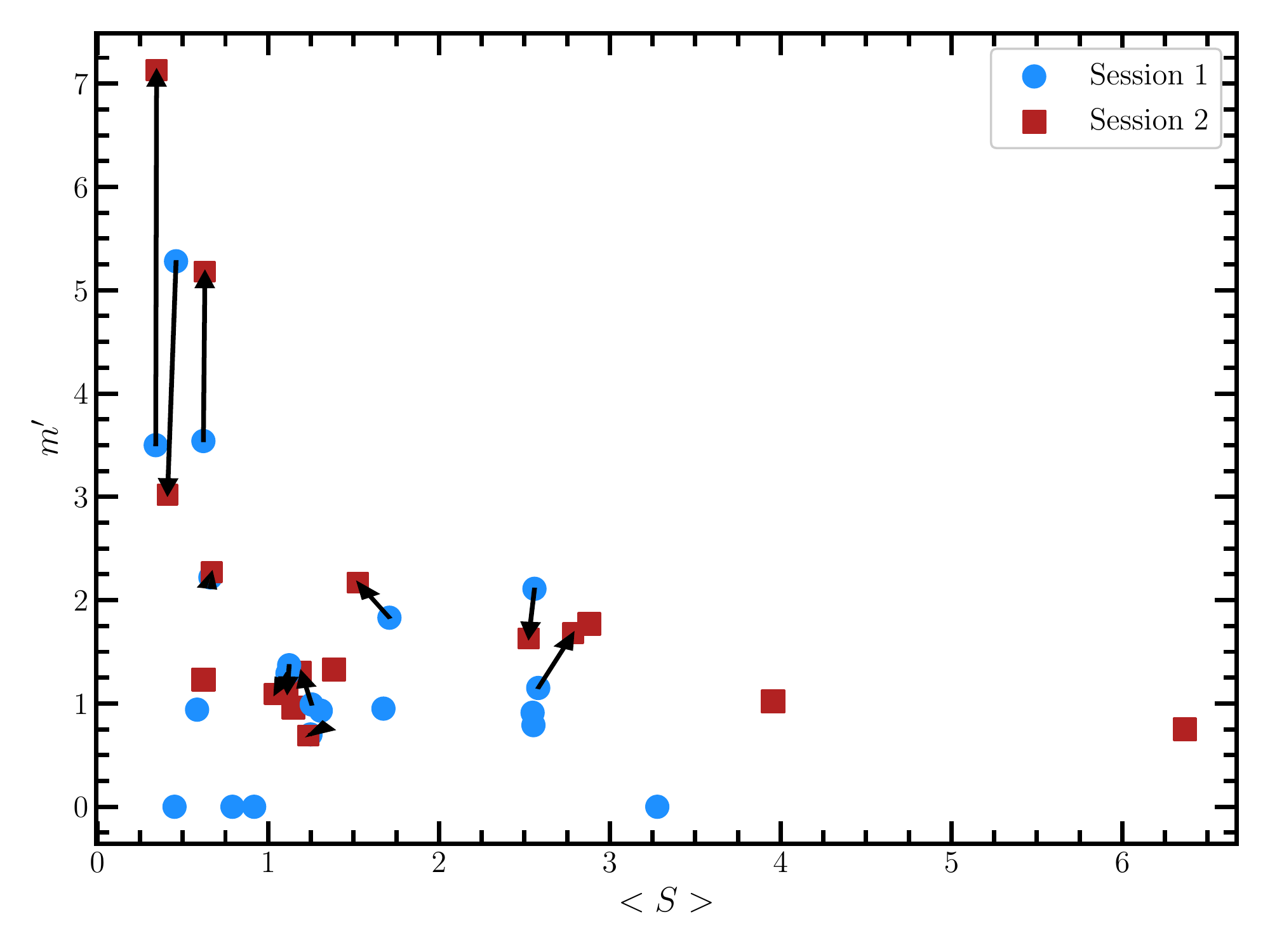}
\caption{The modified modulation index ($m'$ in per cent) versus average flux density ($<S>$, in~Jy) of the target sources observed in session 1 (blue circle) and session 2 (red square). Arrows indicate from session 1 to session 2 for 11 common target~sources.}
\label{fig4}
\end{figure}

\begin{specialtable}[H]
\caption{Results of correlation analysis for modified modulation index ($m'$), redshift ($z$) and Galactic~latitude. 	Column 1: Sample (Full---all target sources in Tables~\ref{tab1} and \ref{tab2}; S1---target sources in Table~\ref{tab1}; S2---target sources in Table~\ref{tab2}). Column 2 and 3: Pearson correlation coefficient and the probability of no correlation. Column 4 and 5: Spearman correlation coefficient and the probability of no correlation.}

\setlength{\cellWidtha}{\columnwidth/5-2\tabcolsep+0.0in}
\setlength{\cellWidthb}{\columnwidth/5-2\tabcolsep+0.0in}
\setlength{\cellWidthc}{\columnwidth/5-2\tabcolsep+0.0in}
\setlength{\cellWidthd}{\columnwidth/5-2\tabcolsep+0.0in}
\setlength{\cellWidthe}{\columnwidth/5-2\tabcolsep+0.0in}
\scalebox{1}[1]{\begin{tabularx}{\columnwidth}{>{\PreserveBackslash\centering}m{\cellWidtha}>{\PreserveBackslash\centering}m{\cellWidthb}>{\PreserveBackslash\centering}m{\cellWidthc}>{\PreserveBackslash\centering}m{\cellWidthd}>{\PreserveBackslash\centering}m{\cellWidthe}}
\toprule
\noalign{\smallskip}
\bf{Sample}	& \bf{Pearson}	& \bm {$p_{null}$}	& \bf {Spearman}	& \bm {$p_{null}$} 	\\
\bf{1}		& \bf{2}		& \bf{3}			& \bf{4}			& \bf{5}			\\
\midrule
\noalign{\smallskip}
\multicolumn{5}{c}{\bf{Modified modulation index ($m'$) versus redshift (z)}}			\\
\noalign{\smallskip}

Full		& $-$0.121		& 0.475				& $-$0.005			& 0.978				\\
S1			& $-$0.173		& 0.465				& $-$0.242			& 0.305				\\
S2			& $-$0.032		& 0.903				& 0.279				& 0.277				\\
\midrule
\noalign{\smallskip}
\multicolumn{5}{c}{\bf{Modified modulation index ($m'$) versus Galactic latitude}}		\\
\noalign{\smallskip}

Full		& 0.016			& 0.926				& $-$0.128			& 0.449		 		\\
S1			& $-$0.045		& 0.850				& $-$0.122			& 0.607				\\
S2			& $-$0.038		& 0.886				& $-$0.272			& 0.290				\\

\bottomrule
\noalign{\smallskip}
\end{tabularx}}
\label{tab3}
\end{specialtable}

For 11 common target blazars in Tables~\ref{tab1} and~\ref{tab2}, there is no significant correlation between the changes of IDV strength $m'$ and the changes of average flux density $<S>$ from session 1 (in February) to session 2 (in March), as~the arrows shown in Figure~\ref{fig4} and the comments in Section~\ref{sec:sources}.

The detection of an IDV or not from one epoch to another is dependent on many factors, e.g.,~the changes of physical properties of scattering plasma as well as the changes of radio core of AGN for the ISS model~\cite{Lovell2008}, and~the telescope sensitivity/uncertainty, calibration procedure/precision, and~also the definition of an IDV detection (slightly different criteria from different research groups). We can see that some of our detections (non-detections) of IDV have non-detections (detections) in previous observations e.g.,~in~\cite{Lovell2008} as shown in Tables~\ref{tab1} and \ref{tab2}, these occur mostly for weak or marginally detected IDVs. The~non-detection of IDV in our session 1 but a weak IDV in session 2 for 1156 + 295 might be partly due to the higher residual uncertainty ($m_0 = 0.64$) of observation in session 1 than ($m_0 = 0.45$) in session 2. Except~for the observation sensitivity (including weather and interferences) effects and the issue of different definition criteria (or confidence level) of IDV, it is believed that the ISS-induced IDVs are often intermittent in time, e.g.,~the strong IDV of 0917 + 624 has been quenched after year 2000~\cite{Liux2015}, although~it shows weak IDV in this~paper.

In the ISS models, the~variability timescale of around a day, requires the angular size of radio-emitting component from sub-milliarcsecond to microarcsecond for the blazar emission to be scintillated by a nearby plasma so as to show an IDV (see~\cite{,Lovell2008,Bignall2006,Koay2011}). Emission from the extended or long jet of a blazar is less possible to show an IDV. However, a~detailed VLBI-structure and core fraction analysis for these blazars along with the brightness temperature of core components from the high-resolution RadioAstron data is planned in the forthcoming~paper.


\section{Summary}

We present the first results of the IDV observations of about two dozens of blazars with Urumqi 25 m radio telescope in 8--12 February 2014 and 7--9 March. Most of target sources are selected in parallel from the observing list of RadioAstron AGN monitoring campaign. The~majority of the blazars exhibit IDV with a high confidence level, some of them show prominent IDVs in the two sessions. Seven new IDV sources are detected in our~observations.

We find that the three strongest IDVs appear in relatively lower flux density $<S> < 0.7$ Jy in the sample, in~which blazar 1357 + 769 is discovered as strong IDV for the first time. No significant correlation between the IDV strength and either redshift or Galactic latitude is found from this blazar~sample.

\vspace{6pt}



\authorcontributions{Conceptualization, writing---original draft, X.L.; analysis and plotting, X.W., N.C.; and review and editing, J.L., L.C., X.Y. and T.P.K. All authors have read and agreed to the published version of the~manuscript.}

\funding{This work is supported by the National Key R\&D Program of China under grant No. 2018YFA0404602, and~partly supported by Xiaofeng Yang's Xinjiang Tianchi Bairen project and CAS Pioneer Hundred Talents Program, and~the CAS `Light of West China' Program under grant No. 2018-XBQNXZ-B-021 and the NSFC under grant No. U2031212 and 61931002. L.C. is also supported by the Youth Innovation Promotion Association of the CAS under grant No.~2017084.}

\institutionalreview{Not applicable.
}

\informedconsent{Not applicable.
%
}

\dataavailability{Not applicable.
}

\acknowledgments{We thank the referees for their helpful comments. This work is based on observations made with the Urumqi 25 m radio Telescope, which is operated by the Key Laboratory of Radio Astronomy, Chinese Academy of Sciences. The~NASA/IPAC Extragalactic Database (NED) is funded by the National Aeronautics and Space Administration and operated by the California Institute of Technology.}

\conflictsofinterest{The authors declare no conflict of~interest.}

\end{paracol}
\reftitle{References}

\end{document}